\documentclass[12pt]{article}

\catcode`\@=11

\global\arraycolsep=2pt
\usepackage{amsbsy,amssymb,latexsym,amsfonts,amsmath}
\usepackage{graphicx,color}

\begin{document}
\begin{flushright}
\parbox{4.2cm}
{CALT-TH-2014-161}
\end{flushright}

\vspace*{0.7cm}

\begin{center}
{ \Large Imaginary supergravity or Virial supergravity?}
\vspace*{1.5cm}\\
{Yu Nakayama}
\end{center}
\vspace*{1.0cm}
\begin{center}
{\it California Institute of Technology,  \\ 
452-48, Pasadena, California 91125, USA}
\vspace{3.8cm}
\end{center}

\begin{abstract}
When a globally supersymmetric theory is scale invariant, it must possess a Virial supercurrent supermultiplet. The multiplet structure is analogous to the R-current supermultiplet in globally R-symmetric theories but we put extra ``$i$"s in various formulae.
We construct a novel type of supergravity  from gauging the Virial supercurrent supermultiplet in $d=1+3$ dimensions. We give the full non-linear superspace action with the help of a covariantly linear  unitary superconformal compensator. The resulting supergravity is peculiar: (1) no Einstein-like second order kinetic term is allowed without matter (2) there exists a dynamical non-geometrical connection (3) the metric is unimodular in the Wess-Zumino gauge and can be coupled  only to scale invariant matter. Examples of matter couplings and higher derivative kinetic terms are studied from the superconformal approach. Our work completes the classification of the irreducible $\mathcal{N}=1$ supergravities in $d=1+3$ dimensions.

\end{abstract}

\thispagestyle{empty} 

\setcounter{page}{0}

\newpage

\section{Introduction}
There are two paths to the (super)gravity. The Venus physicist way or the Earth physicist way \cite{Feynman:1996kb}. The Venus physicist way is to start with the free spin two gauge theory (a.k.a Fierz-Pauli theory) and then try to couple with matter and themselves order by order. This is also called the Noether approach, and it is quite successful in usual spin one gauge theories. Take QED for example. We start with a free $U(1)$ gauge theory and couple it to a free Dirac fermion. The first order interaction is $a^\mu j_\mu$, where $j_\mu = (\bar{\psi}\gamma_\mu \psi)$ is the conserved $U(1)$ current in the free Dirac theory. Then the entire theory is just it! There is no need for the higher corrections although you may add non-minimal couplings such as Fermi terms. Yang-Mills theory is a little bit more complicated because you have to add more terms (such as second order self couplings or contact terms) to ensure the self consistency. Besides, the gauge transformation must be modified from the linearized transformation in the free theory.
Still the modification ends at the second order.

The Venus physicist way is much more complicated in the (super)gravity. It is easy to write down the linearized action and the first order coupling to matter. However, we have to supplement higher and higher order terms to ensure the gauge invariance to retain the physics of the spin two gauge theory. The interaction quickly becomes highly non-linear, and infinitely many terms appear (in terms of the original fields). Nevertheless, the resulting theory has a beautiful geometrical meaning and can be packaged into the compact form of the Einstein-Hilbert action (see e.g. \cite{Deser:1969wk}\cite{Deser:1979zb} for the accomplishment in the Venus physicist way). 

What Einstein did (as a representative of the Earth way) is opposite. He started with the geometry by postulating the full non-linear gauge symmetry (diffeomorphism invariance) at once. Then he wrote down the simplest equations (or action) compatible with the full non-linear gauge symmetry. This was possible because what we observe in our daily lives as gravity is not a propagating spin two gauge field but the geodesic motion realized as an equivalence principle. 
Indeed, since the geometrical meaning has been so much emphasized, it took some time for us to recognize that the Einstein theory of gravity is a theory of spin two gravitons. 

In case of supergravity, we have employed the hybrid of both approaches (see e.g. \cite{Gates:1983nr}\cite{Buchbinder:1998qv} for reviews). 
We may start with the superspace geometry, but the most general superspace geometry is too large to encode the field theories living in $1+3$ dimensional space-time. We need to impose various non-trivial constraints on the superspace geometry to describe the supersymmetric spin two gauge theories coupled to supersymmetric matter and themselves. 

At the linearized level, the spin two gauge field couples to the energy-momentum tensor. In supersymmetric cases, however, the supermultiplet containing the energy-momentum tensor is not unique. Depending on the choice of the supermultiplet that contains the energy-momentum tensor, the Venus physicist approach would be different, and we end up with different formulations of supergravity. From the Earth physicist viewpoints, the constraints on the superspace geometry may not be unique.

The most common choice is the so-called Ferrara-Zumino supermultiplet \cite{Ferrara:1974pz}. The corresponding supergravity is known as the old minimal supergravity \cite{Wess:1978bu}\cite{Stelle:1978ye}\cite{Ferrara:1978em}. It is one of the minimal supergravities in the sense that the number of auxiliary fields (assuming that the kinetic term contains the Einstein-Hilbert term) is minimal (with $12+12$ degrees of freedom). Another example of  minimal supergravity is the new minimal supergravity \cite{Akulov:1976ck}\cite{Sohnius:1981tp}, which is obtained from gauging the R-current supermultiplet that contains not only the energy-momentum tensor but also the R-symmetry current \cite{Gates:1981yc}. The corresponding supergravity is known as the new minimal supergravity, and it can only couple to matter with an R-symmetry. 

There is yet another minimal choice of the supercurrent supermultiplet\footnote{Beyond the minimal choice, there has been renewed interest in various non-minimal supercurrent supermultiplets including the S-multiplet \cite{Magro:2001aj}\cite{Komargodski:2010rb}\cite{Dumitrescu:2011iu}\cite{Kuzenko:2011rd}.} that contains the energy-momentum tensor. When a globally supersymmetric theory is scale invariant, it possesses a Virial supercurrent supermultiplet \cite{Kuzenko:2010am}\cite{Zheng:2010xx}\cite{Kuzenko:2010ni}\cite{Nakayama:2012nd}. The multiplet structure is analogous to the R-current supermultiplet in globally R-symmetric theories but we put extra ``$i$"s in various formulae. The corresponding free spin two gauge theory was constructed in \cite{Gates:2003cz}\cite{Buchbinder:2002gh}, and we would like to address the question what would be the full non-linear supergravity.

It will turn out that the resulting supergravity (called the Virial supergravity) is very peculiar: (1) no Einstein-like second order kinetic term is allowed without matter (2) there exists a dynamical non-geometrical connection (3) the metric is unimodular in the Wess-Zumino gauge and can be coupled  only to scale invariant matter. 
Indeed, the model resembles a certain limit of the non-minimal supergravity  (with $20+20$ degrees of freedom). In the non-minimal supergravity, we have a complex parameter $n$ \cite{Siegel:1978mj}, and our Virial supergravity can be regarded as the $n\to \infty$ limit with an additional constraint (so that the supergravity sector becomes minimal) similarly to that the $n\to 0$ limit with a reality condition corresponds to the new minimal supergravity.

In the literature, we encounter that ``$n\to \infty$ does not lead to a sensible theory" in the non-minimal supergravity (e.g. \cite{Gates:1983nr} among others), which is true in the sense that the Einstein-Hilbert term (without matter) is not allowed in this limit. However, in recent theoretical applications of supergravity, the Einstein-Hilbert term is not necessarily required. In applications to the localization computation of the supersymmetric field theories in curved space-time, the off-shell formulation of the supergravity is essential but there is no need for the Einstein-Hilbert term (or any supergravity dynamics) \cite{Festuccia:2011ws}\cite{Dumitrescu:2012ha}.\footnote{In the same spirit, the $(0,2)$ Virial supergravity was studied in $d=1+1$ dimensions in \cite{Nakayama:2013coa}. In $d=1+1$ dimensions, the Einstein-Hilbert action is topological, so there is no disadvantage at all.} Furthermore, since higher derivative kinetic terms are allowed in the Virial supergravity, they may substitute for the Einstein-Hilbert term as recently argued that with appropriate boundary conditions, conformal gravity may be equivalent to Einstein gravity \cite{Maldacena:2011mk}. Even pure $R^2$ gravity is equivalent to Einstein gravity with a scalar field.

The organization of the paper is as follows. In section 2, we review the Virial  supercurrent supermultiplet and its superspace conservation equations. In section 3, we construct the full non-linear superspace Virial supergravity action with the help of a covariantly linear  unitary superconformal compensator. In section 4, we analyze the linearized action of the Virial supergravity and study its connection to the Virial supercurrent supermultiplet. In section 5, we conclude with  further discussions.

We cannot avoid a certain amount of heavy notation in supergravity. Following the tradition of the community, we do not define all the supergravity nomenclature and conventions within the paper. Our conventions follow mostly \cite{Buchbinder:1998qv} except for some minor changes of fonts and characters. 

\section{Virial supercurrent supermultiplet}

When a supersymmetric theory in $d=1+3$ dimensions is scale invariant, the theory possesses the Virial supercurrent supermultiplet.\footnote{In this section, all superfields are expressed in flat (or rigid/global) $\mathcal{N}=1$ superspace.}
It is defined by the set of  superspace conservation equations
\begin{align}
\bar{D}^{\dot{\alpha}} J_{\alpha \dot{\alpha}}^V &= i{\eta}_{{\alpha}} \  \cr
\bar{D}_{\dot{\alpha}} \eta_{\alpha} &= 0 \cr
D^\alpha \eta_{\alpha} - \bar{D}_{\dot{\alpha}} \bar{\eta}^{\dot{\alpha}} &= 0 \ . \label{defv}
\end{align}
Here $J_{\alpha \dot{\alpha}}^V$ is real and $\eta_{\alpha}$ is chiral.

When we expand $J_{\mu}^J = -\frac{1}{2}J_{\alpha \dot{\alpha}} \bar{\sigma}_\mu^{\dot{\alpha} \alpha}$ and $\eta_{\alpha}$ in components, we find (see e.g. \cite{Dienes:2009td})
\begin{align}
J_{\mu}^V &= j_\mu + \theta^\alpha(S_{\mu\alpha} + \frac{1}{3}\sigma_{\mu \alpha \dot{\alpha}} \bar{\sigma}^{\dot{\alpha} \beta}_\nu S^{\nu}_{\beta} )
+ \bar{\theta}_{\dot{\alpha}}(\bar{S}_\mu^{\dot{\alpha}} + \frac{1}{3}\bar{\sigma}_\mu^{\dot{\alpha} \alpha} \sigma^\nu_{\alpha \dot{\beta}} \bar{S}^{\dot{\beta}}_\nu)+ 2(\theta \sigma^\nu \bar{\theta}) \hat{T}_{\nu\mu} + \cdots \cr
\eta_{\alpha} &= -i\lambda_{\alpha}' + (\delta_{\alpha}^\beta D' - 2i\sigma^\mu_{\alpha \dot{\alpha}} \bar{\sigma}^{\nu \dot{\alpha} \beta} F'_{\mu\nu}) \theta_{\beta} + \theta^2 \sigma^\mu_{\alpha \dot{\alpha}} \partial_\mu {\bar{\lambda}}^{'\dot{\alpha}} + \cdots \ ,
\end{align}
where, due to \eqref{defv}, some of the components are related:
\begin{align}
\hat{T}_{\mu\nu} &= T_{\mu\nu} - \frac{1}{4}F'_{\mu\nu} + \frac{1}{2}\epsilon_{\mu\nu\rho\sigma}(\partial^\rho j^\sigma - \partial^\sigma j^\rho) \cr
D'&= -\partial^\mu j_\mu \cr
\lambda'_{\alpha} &= \frac{1}{3}(\sigma^\mu_{\alpha \dot{\alpha}} \bar{S}^{\dot{\alpha}}_\mu) \ .
\end{align}
The ``energy-momentum tensor" $T_{\mu\nu}$ is conserved $\partial^\nu T_{\mu\nu} = 0$ and traceless $\eta^{\mu\nu} T_{\mu\nu} = 0$, but it is not symmetric: $T_{\mu\nu} -T_{\nu\mu} = \frac{1}{4} F'_{\mu\nu}$, where $F'_{\mu\nu}$ is a closed two-form. Therefore, the Virial supercurrent supermultiplet describes a supersymmetric field theory with manifest translational invariance and scale invariance. The Lorentz invariance is not manifest, however \cite{Dumitrescu:2011iu}. Furthermore neither R-symmetry nor (super)conformal symmetry are necessary.

Since the supercurrent conservation \eqref{defv} demands that the anti-symmetric tensor $F'_{\mu\nu}$ is closed, at least locally we can introduce the potential $B_\mu$ by $-\frac{1}{4}F'_{\mu\nu} = \partial_\mu B_\nu - \partial_\nu B_\mu$. Then we can construct the Belinfante energy-momentum tensor
\begin{align}
\tilde{T}_{\mu\nu} = T_{\mu\nu} + \partial_\mu B_\nu - \eta_{\mu\nu} (\partial_\rho B^\rho) \ 
\end{align}
such that it is symmetric $\tilde{T}_{\mu\nu} = \tilde{T}_{\nu\mu}$ and conserved $\partial^\nu \tilde{T}_{\mu\nu} = 0$. The existence of the symmetric energy-momentum tensor guarantees the Lorentz invariance, but now the trace of the Belinfante energy-momentum tensor does not have to vanish 
\begin{align}
\eta^{\mu\nu} \tilde{T}_{\mu\nu} = -3\partial_\mu B^\mu  \ , \label{tracesym}
\end{align}
which makes it clear that the theory is not necessarily (super)conformal invariant.\footnote{Recall that the conformal invariance requires a symmetric {\it and} traceless energy-momentum tensor. They are not simultaneously realized here. The distinction between scale invariance and conformal invariance is reviewed in \cite{Nakayama:2013is}.} The not-necessarily-conserved current $B_\mu$ is known as the Virial current. 

The superspace statement of the corresponding argument is that we can (locally) write $\eta_{\alpha} = -\frac{1}{2}\bar{D}^2 D_{\alpha} O$ with a real superfield $O$. Then we can (locally) construct the Ferrara-Zumino supermultiplet 
\begin{align}
J_{\alpha \dot{\alpha}}^{\mathrm{FZ}} &= J_{\alpha \dot{\alpha}}^V - i\{D_{\alpha}, \bar{D}_{\dot{\alpha}} \} O \cr
X & = -\frac{i}{2} \bar{D}^2 O  
\end{align}
with the supercurrent conservation equations
\begin{align}
\bar{D}^{\dot{\alpha}} J_{\alpha \dot{\alpha}}^{\mathrm{FZ}} &= D_{\alpha} X \cr\bar{D}_{\dot{\alpha}} X & = 0 \ .
\end{align}
This allows us to move between Ferrara-Zumino supermultiplet and Virial supercurrent  supermultiplet locally, but the global existence of $O$ may become an obstruction.\footnote{An example is a free gauge invariant two-form tensor theory contained in a linear real multiplet, where we can define a gauge invariant Ferrara-Zumino supermultiplet, but the Virial supercurrent supermultiplet is not gauge invariant \cite{Nakayama:2012nd}.}

Given a scale invariant supersymmetric field theory, the Virial supercurrent supermultiplet may not be unique. It admits the superimprovement \cite{Nakayama:2012nd}. If the theory possesses a (non-R) conserved current $j^f_\mu$, we have the corresponding real superfield $J^f$ with $D^2 J^f = \bar{D}^2 J^f = 0$. Then we can always construct the improved Virial supercurrent supermultiplet as
\begin{align}
J^V_{\alpha \dot{\alpha}} &\to J_{\alpha \dot{\alpha}}^V + i\{ D_{\alpha}, \bar{D}_{\dot{\alpha}}\} J^f \cr
\eta_{\alpha} &\to \eta_{\alpha} - \frac{1}{2}\bar{D}^2 D_{\alpha} J^f \ . 
\end{align}
We can easily check that the supercurrent conservation equations \eqref{defv} are intact.

In components, the most important effect of the superimprovement is that we add the conserved current $j^f_{\mu}$ to the antisymmetric tensor field $F'_{\mu\nu}$ by its rotation $\partial_\mu j^f_\nu - \partial_\nu j^f_\mu$ and consequently the (non-symmetric) energy-momentum tensor $T_{\mu\nu}$ is shifted by $\partial_\mu j^f_\nu$ without affecting the conservation (because $\partial^\mu j^f_\mu = 0$). This in turn means that the Virial current $B_\mu$ acquires the extra contribution from the conserved current $j^f_\mu$. Since it is conserved, the trace of the improved Belinfante energy-momentum tensor $\tilde{T}_{\mu\nu}$ in  \eqref{tracesym} does not change.

Suppose we would like to gauge the Virial supercurrent supermultiplet. The gauge field will be the real (linearized) vierbein superfield  $H_{\alpha \dot{\alpha}}$.
We postulate the linearized coupling
\begin{align}
\int d^4x d^4\theta J^V_{\alpha \dot{\alpha}} H^{\alpha \dot{\alpha}} \label{linearizedc}
\end{align}
with the gauge transformation
\begin{align}
\delta H_{\alpha \dot{\alpha}} = D_{\alpha} \bar{L}_{\dot{\alpha}} - \bar{D}_{\dot{\alpha}} L_{\alpha} \ , \label{flatgauge}
\end{align}
where $L_{\alpha}$ is a spinor gauge parameter.
In order for the linearized coupling \eqref{linearizedc} to be invariant, the supercurrent conservation equations  \eqref{defv} demand that the gauge parameter $L_{\alpha}$ must be constrained as
\begin{align}
\bar{D}_{\dot{\alpha}}D^2 \bar{L}^{\dot{\alpha}} + D_{\alpha} \bar{D}^2 L^\alpha = 0 \ . \label{constv}
\end{align}

It is sometime cumbersome to deal with the constrained gauge transformation. The constraint can be easily avoided by introducing the compensator and enlarging the gauge symmetry. In our case, we avoid the constraint \eqref{constv} by introducing the spinor chiral superfield $\psi_{\alpha}$  ($\bar{D}_\alpha \psi_\beta = 0$) as a compensator with the gauge transformation
\begin{align}
\delta \psi_{\alpha} = i \bar{D}^2 L_\alpha \ . \label{spinorcom}
\end{align}

The invariant action under the unconstrained gauge transformation is
\begin{align}
\int d^4x d^4 \theta  J^V_{\alpha \dot{\alpha}} H^{\alpha \dot{\alpha}}  + \int d^4x d^2\theta \psi^{\alpha} \eta_\alpha + \int d^4x d^2\bar{\theta} \bar{\psi}_{\dot{\alpha}} \bar{\eta}^{\dot{\alpha}} \ . \label{act}
\end{align}
The unconstrained gauge transformation contains the linearized super Weyl transformation, so the compensator $\psi_{\alpha}$ is also called the superconformal compensator. Out of $\psi_{\alpha}$, one can construct the linear real superconformal compensator 
\begin{align}
 U = \frac{1}{12} \left(D^{\alpha}\psi_\alpha + \bar{D}_{\dot{\alpha}} \bar{\psi}^{\dot{\alpha}} \right)
\end{align}
such that $\bar{D}^2 U = D^2 U= 0$. It is equipped with the gauge transformation
\begin{align}
\delta U = \frac{i}{12}(D^\alpha \bar{D}^2 L_\alpha - \bar{D}_{\dot{\alpha}} D^2 \bar{L}^{\dot{\alpha}} ) \ .\label{utrans}
\end{align}
Since we saw that $\eta_{\alpha}$ can be locally expressed as $\eta_{\alpha} = -\frac{1}{2}\bar{D}^2 D_{\alpha} O$, we may rewrite the superpotential term in \eqref{act} as
\begin{align}
\int d^4x d^2\theta \psi^{\alpha} \eta_\alpha + \int d^4x d^2 \bar{\theta} \bar{\psi}_{\dot{\alpha}} \bar{\eta}^{\dot{\alpha}} = 6 \int d^4x d^4\theta UO \ .
\label{act} \end{align}  

We note that in general $O$ (therefore the corresponding  Ferrara-Zumino supermultiplet) is ambiguous under the ``superimprovement" $O \to O + \Omega + \bar{\Omega}$, where $\Omega$ is a chiral superfield. Since $O$ couples to a linear real superfield $U$, this ambiguity does not change the action \eqref{act} after the superspace integration. Within this class of gauge transformation, the local nature of the existence of $O$ does not cause problems in gauging. Our discussion here is analogous to the gauging of an R-current of $U(1)$ gauge theories with Fayet-Iliopoulos parameters, where the Ferrara-Zumino supermultiplet is not gauge invariant, but the coupling in the new minimal supergravity is still possible.

In \cite{Buchbinder:2002gh}\cite{Gates:2003cz}\, the gauge invariant second order kinetic term for $H_{\alpha\dot{\alpha}}$ and $U$ was constructed. For completeness, we quote the result here
\begin{align}
S_{\mathrm{free}} = \int d^4x d^4\theta \left(H^{\alpha \dot{\alpha}} (\Box \Pi) H_{\alpha \dot{\alpha}}  + U \partial_{\alpha \dot{\alpha}} H^{\alpha \dot{\alpha}} + \frac{3}{2}U^2 \right) \ ,\label{laction}
\end{align}
where the projector $\Pi$ is given by
\begin{align}
\Pi H_{\alpha \dot{\alpha}} &= \frac{1}{3}\Pi^L_{\frac{1}{2}} H_{\alpha \dot{\alpha}} + \frac{1}{2}\Pi^{T}_{\frac{3}{2}} H_{\alpha \dot{\alpha}} \cr
& = \frac{1}{48}\Box^{-2} \partial_{\alpha \dot{\alpha}} D^\gamma \bar{D}^2 D_{\gamma} \partial_{\beta \dot{\beta}} H^{\beta \dot{\beta}} -\frac{1}{96} \Box^{-2} \partial_{\dot{\alpha}}^\beta D^\gamma \bar{D}^2 D_{(\gamma} \partial_{\alpha}^{\dot{\beta}} H_{\beta)\dot{\beta}} \ . 
\end{align}
In the Wess-Zumino gauge, the action \eqref{laction} describes the propagation of a supersymmetric massless spin two particle with the linearized Einstein-Hilbert term where the linearized trace mode of the metric is dualized to a two-form tensor.

Our goal is to obtain the full non-linear extension of the above gauging, which should lead to a novel formulation of the supergravity based on the Virial supercurrent supermultiplet. It could be instructive to continue this Noether approach (or ``Venus physicists approach") step by step particularly because it would {\it fail}. As we will see, the naive extension of the second order kinetic term for $H_{\alpha\dot{\alpha}}$ and $U$   \eqref{laction} would not exist beyond the linearized level (without additional matter).\footnote{Apparently, this has been recognized by supergravity experts. The author would like to thank S.~Kuzenko and  W.~D.~Linch for discussions.} Instead we will pursue the structure in  the full non-linear theory (to be called Virial supergravity) from the superconformal approach directly in the curved superspace. Then we linearize the theory around flat superspace to see its connection to the Virial supercurrent supermultiplet.

\section{Virial supergravity in superspace}

Our construction of the Virial supergravity, which is obtained by gauging the Virial supercurrent supermultiplet, is based on the superconformal compensator approach in superspace. We start with the super Weyl invariant action given in a conventional supergravity (say old minimal supergravity) with a superconformal compensator $\Psi$ (which is tensor type in the language of \cite{Gates:1983nr}). Then we fix the gauge symmetry under the super Weyl transformation which makes the tensor type compensator into the density type compensator. Depending on the choice of the compensator (and of course the super Weyl invariant action), the resulting supergravity theory is different (but sometimes equivalent). For example, the choice of $\Psi$ as a covariantly chiral superfield leads to the old minimal supergravity, and the choice of $\Psi$ as a  covariantly linear real superfield leads to the new minimal supergravity.

Our choice of $\Psi$ is a covariantly linear {\it unitary} superfield:
\begin{align}
(\bar{\mathcal{D}}^2 -4R) V &= 0  \cr
V \bar{V} &= 1 \ . \label{covuni}
\end{align}
These conditions are consistent with the super Weyl transformation
\begin{align}
\varphi \to \varphi' &= e^{\sigma} \varphi \cr
 V \to V' & = e^{\sigma - \bar{\sigma}} V 
\end{align}
with an arbitrary covariatnly chiral superfield $\sigma$: $\bar{\mathcal{D}}_{\dot{\alpha}} \sigma  = 0$. Note in particular that only this super Weyl weight for $V$ is consistent with the covariantly linear unitary constraint \eqref{covuni}. 
Without the unitarity constraint $V\bar{V} =1$, the set of constraints would be  the same as those in the non-minimal supergravity in the $n\to \infty$ limit. In the non-minimal supergravity, the covariantly linear {\it complex} compensator transforms as $\Psi \to \Psi' = \exp(\frac{3n-1}{3n+1}\sigma - \bar{\sigma}) \Psi$ under the super Weyl transformation.
Similarly to the case in the $n=0$ limit, where we can impose the reality condition  on $\Psi$ (leading to the new minimal supergravity), the covariantly linear constraint is reducible in this opposite limit $n\to\infty$, and we have introduced the unitarity constraint.
We have introduced the density type compensator $\varphi$ of the old minimal supergravity  (in which $n=-\frac{1}{3}$)  here in the definition of the super Weyl transformation, and we will further employ it to construct the super Weyl invariant action, but any other formalism is fine. 

Once we determined the superconformal compensator, the rest of the construction of the supergravity is straightforward in theory. We write down the super Weyl invariant action with the compensator (coupled to extra matter if necessary) and fix the superconformal gauge $\varphi = 1$, then the tensor type compensator $\Psi$ (in our case $V$) becomes a new density type compensator for the resulting supergravity action. 

Let us start with the pure supergravity action. With the covariantly linear compensator $\Psi$ in the non-minimal supergravity, the simplest super Weyl invariant action was 
\begin{align}
 S_{\mathrm{non-minimal}} = \frac{1}{n\kappa^2} \int d^8z E^{-1} (\bar{\Psi}\Psi)^{\frac{3n+1}{2}} \ . \label{nonminimals}
\end{align}
Indeed, after fixing the superconformal gauge $\varphi=1$, the resulting on-shell theory is equivalent to the old minimal Einstein supergravity (when $n\neq 0,-\frac{1}{3},\infty$) with different sets of auxiliary fields \cite{Breitenlohner:1977jn}\cite{Brink:1978iv}. 

In the new minimal supergravity (i.e. $n=0$), we impose the reality condition $\Psi = L = \bar{L}$ on the covariantly linear compensator, and we take the $n\to 0 $ limit of \eqref{nonminimals} carefully, resulting in the new minimal supergravity action
\begin{align}
 S_{\mathrm{new-minimal}} = \frac{3}{\kappa^2} \int d^8z E^{-1} L (\log{L} -1) \  \label{newminimals} \ .
\end{align}
Again the on-shell theory is equivalent to the old minimal Einstein supergravity but with a different set of auxiliary fields.

However, in our Virial supergravity (i.e. $n=\infty$), the naive replacement of $\Psi$ with the linear unitary compensator $V$ does not work because we have the unitarity condition $V\bar{V} = 1$ and the candidate action \eqref{nonminimals} with $\bar{\Psi}\Psi = 1$ is not super Weyl invariant for finite $n$ and vanishes in the $n\to \infty$ limit. 
The pure Virial supergravity therefore does not admit the Einstein-like second order derivative kinetic term. This is in contrast to the linearized spin two theory with the superspace action \eqref{laction}, where the second order kinetic term is allowed and it is duality equivalent  to the other formulations \cite{Gates:2003cz}. The failure of the Noether procedure may be traced back to the fact that Einstein gravity is not scale invariant while the linearized theory is. We will discuss the underlying symmetry of the Virial supergarvity in section 4.

Actually, there are two other (and as far as we know only two other\footnote{We assume the power series expansion.}) super Weyl invariant actions that can be constructed out of our covariantly linear unitary compensator $V$ (without extra matter). They are
\begin{align}
 S_{\mathrm{Virial}} = \int d^8z \frac{E^{-1}}{R}\left(\alpha \mathcal{W}^{\alpha\beta\gamma} \mathcal{W}_{\alpha\beta\gamma} + \beta \mathcal{W}^{\alpha} \mathcal{W}_{\alpha}\right) + c.c. \ , \label{virihh}
\end{align}
where $\mathcal{W}_{\alpha\beta\gamma}$ is the Weyl tensor chiral superfield that does not depend on $V$, and $\mathcal{W}_{\alpha} = (\bar{\mathcal{D}}^2-4R) \mathcal{D}_{\alpha} \log V$ is the ``field strength superfield" constructed out of $\log V$: The action of the super Weyl transformation has similarity to the supergauge transformation of the ``gauge potential vector superfield" $\log V$.

The first term in \eqref{virihh} does not depend on the compensator, and it is the well-known action for the conformal supergravity (e.g. \cite{Fradkin:1985am} for a review). It contains the Weyl tensor squared term  together with  the conformal invariant vector and spin 3/2 actions. In the new minimal supergravity, we could construct a term similar to the second one (by replacing $V$ with the linear  real compensator), and it contains the $R^2$ term \cite{Cecotti:1987qe}. We will discuss the component form of the second term in the Virial supergravity in the next section.

Although the Einstein-like second order derivative kinetic term is not allowed in the pure Virial supergravity, we may still look for an effective kinetic term from the matter couplings. The matter couplings are also important in addressing the connection to the Virial supercurrent supermultiplet we have discussed in section 2 as our starting point. Since the possible choice of matter couplings are numerous, however,  we only focus on several interesting examples here. 

The first example is the coupling to a covariantly chiral superfield $\chi$ (i.e. $\bar{\mathcal{D}}_{\dot{\alpha}} \chi =0$) with a specific {\it complex} super Weyl weight
\begin{align}
\varphi \to \varphi' &= e^{\sigma} \varphi \cr
\chi \to \chi' & = e^{(-1+i\alpha)\sigma} \chi
\end{align}
where $\bar{\alpha} = \alpha$ is a real parameter which will be related to the improvement of the Virial current (or dilatation current).
The simplest super Weyl invariant action is given by
\begin{align}
\int d^8z E^{-1} \bar{\chi} \chi (V^{-i\alpha} + \bar{V}^{+i\alpha}) \ . \label{chiralW}
\end{align} 
Obviously when $\alpha = 0$, the chiral scalar multiplet is super Weyl invariant, so there is no coupling to the compensator. The non-zero value of $\alpha$ introduces the non-minimal coupling to the Virial supergravity multiplet. At the linearized level, it is related to the improvement of the Virial supercurrent supermultiplet by adding the conserved $U(1)$ current $J = \bar{\chi} \chi$ (with $D^2 J = \bar{D}^2 J = 0$ in the global limit) to the superconformal supercurrent supermultiplet of $\chi$ (when $\alpha = 0$) as we will see in the next section.

Alternatively, we may realize $\chi$ as a chiral superconformal compensator. Then the action may be regarded as the old minimal supergravity (with the chiral compensator $\chi$) coupled to the linear unitary matter multiplet $V$ rather than the Virial supergravity coupled to the chiral matter multiplet $\chi$. From this viewpoint, if we expand the linear unitary multiplet $V$ around unity $V = 1+i\Upsilon + \cdots$ (with $\Upsilon=\bar{\Upsilon}$), the Einstein-Hilbert term does appear from the old minimal supergravity term $\int d^8z E^{-1}$ in \eqref{chiralW} (after the superconformal gauge fixing).

The Virial supergravity only couples to scale invariant theories, but these scale invariant theories do not have to be R-symmetric. Let us take an example of (classically) scale invariant but non R-symmetric (nor superconformal) action mentioned in \cite{Nakayama:2012nd} and try to couple it to the Virial supergravity. Suppose the chiral superfield $\chi$ has super Weyl weight $-\frac{1}{2}$ (i.e. $\chi \to \chi' = e^{-\frac{1}{2}\sigma} \chi)$.  With the use of our linear unitary compensator $V$, we can construct the super Weyl invariant action
\begin{align}
\int d^8z E^{-1}\left( a V^{\frac{1}{2}}\bar{\chi} \chi^3+ b \chi^2 \bar{\chi}^2 + \bar{a} V^{-\frac{1}{2}} \chi\bar{\chi}^3 + \lambda \frac{\chi^6}{R} +  \bar{\lambda} \frac{\bar{\chi}^6}{\bar{R}}\  \right) . 
\end{align}
Note that this theory cannot be coupled to the new minimal supergravity (unless the superpotential term with  coefficient $\lambda$ vanishes) because if we set $V=1$, the theory is still scale invariant but not R-symmetric. 

The second example is the coupling to a covariantly linear real superfield $L = \bar{L}$ (i.e. ($\bar{\mathcal{D}}^2-4R) L = (\mathcal{D}^2-4\bar{R})L = 0$)  with the super Weyl weight
\begin{align}
\varphi \to \varphi' &= e^{\sigma} \varphi \cr
L \to L' & = e^{-\sigma-\bar{\sigma}} L
\end{align}
In this case, it is harder to find the lowest derivative coupling to the Virial supergravity. This is related to the fact that at the lowest derivative order the linear multiplet $L$ describes a gauge invariant two-form tensor field and the simplest free gauge invariant two-form tensor theory with the two derivative kinetic term is not scale invariant.\footnote{Nevertheless the free theory is scale invariant by breaking the gauge invariance in the dilatation current. See e.g. \cite{Nakayama:2012nd}.} One possible choice of the super Weyl invariant action would be
\begin{align}
\int d^8z E^{-1} L (\alpha \log L + \beta \log V) \ , \label{linearaction}
\end{align}
which contains the scalar field dependent  kinetic term for the two-form tensor field in $L$. This scalar dependent kinetic term ensures the gauge invariant Virial current.

Actually, if we regard $L$ as a linear real superconformal compensator, we may realize that the action \eqref{linearaction} is equivalent to the new-minimal supergravity coupled to the linear unitary matter multiplet $V$ rather than the Virial supergravity coupled to the linear real multiplet $L$. In the latter perspective, Einstein action is encoded as the new minimal supergravity kinetic term $\int d^8z E^{-1} L (\alpha \log L)$. With this alternative superconformal gauge fixing, the two-form field in $L$ becomes auxiliary in the Wess-Zumino gauge.

As the third example, let us consider the coupling to a vector multiplet.
It is well-known that the gauge invariant kinetic term for a vector multiplet is super Weyl invariant, so it can be introduced into the Virial supergravity without the direct coupling to the superconformal compensator $V$. The vector multiplet is described by a real superfield $A = \bar{A}$ with the super Weyl weight zero. The gauge invariant kinetic term is given by
\begin{align}
 S_{\mathrm{Maxwell}} = \int d^8z \frac{E^{-1}}{R}\left(\tau W^\alpha W_{\alpha} + c.c.  \right) \ , \label{maxwell}
\end{align}
where  $W_{\alpha} = (\bar{\mathcal{D}}^2-4R) \mathcal{D}_{\alpha} A$ is the covariantly chiral field strength superfield as usual. In components, the action \eqref{maxwell} contains the usual Maxwell action for the vector field $a_\mu$.

The more non-trivial term is the supersymmetric extension of the ``gauge fixing term" $(D^\mu a_{\mu})^2$ that is scale invariant but cannot be made conformal invariant (without a superconformal compensator). With the help of the superconformal compensator $V$,  we may introduce the super Weyl invariant (but not supergauge invariant) term such as 
\begin{align}
\int d^8z E^{-1}\left( (\mathcal{D}^2-4\bar{R})(A\bar{V}) (\bar{\mathcal{D}}^2-4R)(A{V}) \right ) \ . 
\end{align}
This is the superconformal generalization of the ``gauge fixing term" with our covariantly linear unitary compensator $V$. The physical realization of the scale invariant but non-conformal field theories of this type was discussed in the literature \cite{Riva:2005gd}\cite{ElShowk:2011gz}. Unlike the above two examples, we cannot regard the vector superfield $A$ as a superconformal compensator because its Weyl weight vanishes. Coincidentally, the action does not contain the Einstein-like kinetic term for gravity. It would be interesting to compare our examples with the $(16+16)$ reducible supergravity studied in \cite{Gates:2003cz} at the linearized level. In particular, it would be intriguing to see whether we could take the decoupling limit of matter to isolate the effective massless spin two action \eqref{laction}.

\section{Linearization and bosonic component action}
In order to understand the underlying structure of the Virial supergravity, we would like to study the component expression of the Virial supergravity (with various matter couplings) at the linearized level. This will also enable us to connect the properties of the Virial supercurrent supermultiplet reviewed in section 2.

We recall that the Virial supergravity is characterized by the covariantly linear unitary compensator $V$
\begin{align}
(\bar{\mathcal{D}}^2 -4R) V &= 0  \cr
V \bar{V} &= 1 \ .
\end{align}
We may solve the first linear constraint by introducing the flat complex linear superfield ${\gamma}$ and the supergravity prepotential $F$ (as well as $W = W^M\partial_M$),\footnote{$a,b \cdots$ refer to the Lorentz indices while $\mu,\nu \cdots$ refer to the Einstein indices. $A,B \cdots$ denote the super Lorentz indices while $M,N \cdots$ denote the super Einstein indices.}
 as
\begin{align}
 V &= \bar{F}^{-2} {\gamma} \cr 
\bar{D}^2 {\gamma} &= 0 \ .  \label{constr}
\end{align}
See \cite{Buchbinder:1998qv} for more details. 
The prepotential $F$ can be expressed with the help of the old minimal (density type) compensator $\varphi$ as
\begin{align}
F = \varphi^{\frac{1}{2}} \bar{\varphi}^{-1}(1\cdot e^{\overleftarrow{W}})^{-\frac{1}{3}} (1\cdot e^{\overleftarrow{\bar{W}}})^{\frac{1}{6}} \hat{E}^{-\frac{1}{6}} \ .
\end{align}
Here $\hat{E} = \mathrm{sdet}(\hat{E}_{A}^{\ M})$ constructed out of the super vierbein as $E_{\alpha}^M \partial_M = F\hat{E}_{\alpha}^M \partial_M$ and $\hat{E}_a^M \partial_M = -\frac{i}{4} (\bar{\sigma}_a)^{\alpha \dot{\alpha}} \{\hat{E}_{\alpha}, \bar{\hat{E}}_{\dot{\alpha}}\}$.
The unitarity condition $V \bar{V} = 1$ therefore reads
\begin{align}
\gamma \bar{\gamma} =(1\cdot e^{\overleftarrow{W}})^{-\frac{1}{3}} (1\cdot e^{\overleftarrow{\bar{W}}})^{\frac{1}{3}} \hat{E}^{-\frac{2}{3}} \label{constr2}
\end{align}
in the superconformal gauge $\varphi = 1$.

In summary, in the Virial supergravity, we are equipped with the flat linear (density type) compensator ${\gamma}$ with the constraint \eqref{constr} or \eqref{constr2}, which is intrinsic to the Virial supergravity after setting $\varphi = 1$. The remnant of the old minimal supergravity with the chiral compensator $\varphi$ is gone at this point.

We would like to solve these constraints at the linearized level. For this purpose, it is facilitating to use the chiral representation (again see \cite{Buchbinder:1998qv} for more details). 
We first shift the preptential $W$ around the flat superspace 
\begin{align}
e^W \to e^W e^{-i \mathcal{H}_0}
\end{align}
with $\mathcal{H}_0 = \theta \sigma^a \bar{\theta} \partial_a $ so that $W=0$ corresponds to the super Poincar\'e invariant space-time. 
Then we fix the gauge transformation with respect to $\Lambda_{\dot{\alpha}}$ and $\Lambda_{\alpha\dot{\alpha}}$ by demanding the supergravity gauge
\begin{align}
iW + i(\mathrm{Lorentz}) = H^A D_A + (\mathrm{Lorentz}) = H =  H^a \partial_a \ ,
\end{align}
where $(\mathrm{Lorentz})$ is a certain Lorentz transformation that can be gauged away. 
Here, $H_{a}$ can be regarded as the real vierbein superfield.
Within this gauge, the remaining gauge transformations are
\begin{align}
\delta e^{-2iH} &= \Lambda e^{-2iH} -e^{-2iH} \bar{\Lambda} \cr
\delta \gamma & = \Lambda^a\partial_a\gamma + \Lambda^{\alpha}D_{\alpha}\gamma + \bar{D}_{\dot{\alpha}}(\Lambda^{\dot{\alpha}}\gamma) + \frac{1}{3}(\partial_a\Lambda^a - D_{\alpha}\Lambda^{\alpha})\gamma \ , \label{remainingg}
\end{align}
where $\Lambda = \Lambda^A D_A$. The gauge parameters $\Lambda$ in the supergravity gauge are parameterized by
\begin{align}
\Lambda_{\alpha} &= -\frac{1}{4}\bar{D}^2 L_{\alpha} \cr
\Lambda_{\alpha \dot{\alpha}} &= -2i \bar{D}_{\dot{\alpha}} L_{\alpha} \cr
\Lambda_{\dot{\alpha}} &= -\frac{1}{4}e^{-2iH} D^2 \bar{L}_{\dot{\alpha}} \ .
\end{align}
Note that $\Lambda_{\alpha}$ was chiral before imposing the supergravity gauge and the arbitrary spinor superfield parameter $L_{\alpha}$  corresponds to the linearized gauge transformations in \eqref{flatgauge} without constraints.

We now study the Virial supergravity constraints at the linearized level.
We expand $\gamma = 1+\Gamma$ and we keep all the superfields at the first order with respect to $\Gamma$ and the linearized vierbein superfield $H_a$.
The linear constraint on $\gamma$ is simply
\begin{align}
\bar{D}^2 \Gamma = 0 \ ,
\end{align}
so $\Gamma$ is again a flat linear superfield.

The next step is to study the unitarity constraint. For this purpose, we recall the series expansions of prepotentials in the chiral representation  \cite{Buchbinder:1998qv} 
\begin{align}
\hat{E}^{-\frac{1}{3}} &= 1 -\frac{1}{6} \bar{D}_{\dot{\alpha}} D_{\alpha} H^{\alpha \dot{\alpha}} + \cdots \cr
(1\cdot e^{-i\overleftarrow{H}})^{-\frac{1}{3}} &= 1 +\frac{i}{3} \partial_a H^a + \cdots \ , 
\end{align}
and the linearized unitarity constraint becomes 
\begin{align}
\Gamma + \bar{\Gamma} = -\frac{1}{6} \bar{D}_{\dot{\alpha}} D_{\alpha} H^{\alpha \dot{\alpha}} + \frac{1}{6}D_\alpha \bar{D}_{\dot{\alpha}}H^{\alpha \dot{\alpha}}  \ .
\end{align}
Under the linearized gauge transformation, they transform as 
\begin{align}
\delta H_{\alpha \dot{\alpha}} &= \bar{D}_{\dot{\alpha}}L_\alpha - D_\alpha \bar{L}_{\dot{\alpha}} \cr
\delta \Gamma &= -\frac{1}{12} \bar{D}^2 D^\alpha L_{\alpha} + \frac{1}{4}\bar{D}^{\dot{\alpha}} D^2 \bar{L}_{\dot{\alpha}}
\end{align}
with an arbitrary spinor superfield $L_{\alpha}$ (see \eqref{remainingg}).

This allows us to define the linear real compensator 
\begin{align}
U = i (\Gamma +\frac{1}{6}\bar{D}_{\dot{\alpha}} D_{\alpha} H^{\alpha \dot{\alpha}} ) \ 
\end{align}
such that $U = \bar{U}$ and $\bar{D}^2 U = 0$. Under the gauge transformation it transforms as 
\begin{align}
\delta U = \frac{i}{12}(D^\alpha \bar{D}^2 L_\alpha - \bar{D}_{\dot{\alpha}} D^2 \bar{L}^{\dot{\alpha}} ) \ .
\end{align}
This is precisely the compensator we find in the linearized spin two theories studied in \cite{Gates:2003cz}\cite{Buchbinder:2002gh} and reviewed in section 2 (see \eqref{utrans}).

To go further, we take the Wess-Zumino gauge, in which the vierbein superfield takes the form \cite{Gates:2003cz}
\begin{align}
H_{\alpha \dot{\alpha}} = (\theta {\sigma}^\mu \bar{\theta})(h_{\mu\nu} \sigma^{\nu}_{\alpha \dot{\alpha}}) + \theta_\alpha \bar{\theta}_{\dot{\alpha}} h + \theta^2 \bar{\theta}^2 A_{\alpha \dot{\alpha}}  + \mathrm{fermions} \  \label{wzgauge}
\end{align}
with the residual gauge parameters
\begin{align}
L_\alpha = i\bar{\theta}^{\dot{\alpha}} \zeta_{\alpha \dot{\alpha}} + \bar{\theta}^2 \theta_{\alpha} (f+ ig) + \mathrm{fermions} \ . \label{resg}
\end{align}
Here the linearized metric is decomposed into the traceless mode $h_{\mu\nu}$ and the trace mode $h$.
The gauge transformation by a real vector parameter $\zeta_{\alpha \dot{\alpha}}$ can be regarded as the linearized diffeomorphism. The real scalar parameter $f$ will generate the local Weyl transformation while $g$ can be regarded as the (compensated) R symmetry.

Since the gauge degrees of freedom of $g$ is regarded as the compensated R-symmetry transformation, it can be used to gauge away the R-symmetry compensator appearing in the $\theta$ independent components of $U$ (which is originated from the  phase factor in $V$). 
In this gauge, the linearized compensator takes the form \cite{Buchbinder:2002gh}
\begin{align}
U = \frac{1}{3}(\theta \sigma^\mu \bar{\theta}) G_\mu + \mathrm{fermions} \ . 
\end{align} 
The reality condition (which is originated from the unitarity condition on $V$) demands $G_\mu$ is real and divergence free: $\partial^\mu G_\mu =0$, so it may be regarded as a dual field strength of the two-form tensor field (i.e. $G^\mu = \frac{1}{2} \epsilon^{\mu\nu\rho\sigma} \partial_\nu B_{\rho\sigma} $). The potential two-form $B_{\mu\nu}$ with the additional gauge symmetry $\delta B_{\mu\nu} = \partial_\mu \lambda_\nu - \partial_\nu \lambda_\mu$ is naturally contained in the unconstrained but redundant spinor superfield $\psi_{\alpha}$ in \eqref{spinorcom}.

If we were working in the new minimal supergravity, we would still have had the extra (gauged) R-symmetry in the Wess-Zumino gauge. In a similar way, we still have the local Weyl symmetry given by the parameter $f$ which has not been fixed. There is no tensor type compensator for the local Weyl symmetry to fix it in a fully generally covariant way (without matter). We may want to retain it (as in new minimal supergravity), or we may fix it, in the spirit of the Wess-Zumino gauge, by demanding the unimodular condition on the metric (i.e. $\det g = -1$). At the linearized level, it is equivalent to demanding  that the trace mode $h$ vanish in \eqref{wzgauge}.\footnote{This is similar to what happened in the non-minimal supergravity with $n=-\frac{1}{2}$, where the general covariance is apparently lost by taking the unimodular gravity gauge \cite{Siegel:1978mj}.}

After taking this unimodular gravity gauge, the remaining gauge transformation from the parameter $\zeta_\mu$ in \eqref{resg} is the Weyl compensated volume preserving diffeomorphism. Whenever we do the diffeomorphism that changes the volume (i.e. $\partial^\mu\zeta_\mu \neq 0$), we have to perform the simultaneous Weyl transformation to keep the volume element fixed. 
 One point to be noticed, however, is that the vector field $G_\mu$ in $U$ also transforms non-homogeneously under the Weyl compensated volume preserving diffeomorphism \cite{Gates:2003cz} as 
\begin{align}
\delta G_\mu = -\partial^\nu \partial_\nu \zeta_\mu + \partial^\nu \partial_\mu \zeta_\nu \ .
\end{align}
It means that $G_\mu$ is not a geometric tensor. It rather transforms as a connection $-G^\mu \sim \Gamma^\mu = g^{\rho\sigma} \Gamma^\mu_{\ \rho\sigma}$ under the Weyl compensated volume preserving diffeomorphism. Unlike the Christoffel connection $\Gamma_{\mu}$, however, $G_\mu$ is not made out of vierbeins, but is an independent field. It would be called the Virial connection.\footnote{The torsion part of the connection does not matter as long as it is a Weyl invariant tensor.}

As a consequence, the resulting supergravity theory is not diffeomorphism invariant in a usual sense as we will shortly see in examples. This is expected because the natural energy-momentum tensor encoded in the Virial supercurrent supermultiplet is not manifestly symmetric, so the theory cannot be diffeomorphism invariant. Nevertheless, we do possess the invariance under the linearized gauge transformation with respect to the vector parameter $\zeta_\mu$. The non-invariance under the geometric transformation is cancelled by the extra non-geometric transformation of $G_\mu$. From the viewpoint of the current conservation, this is nothing but the effect of the anti-symmetric part of the conserved energy-momentum tensor.

It is worthwhile emphasizing the distinction between our gravity with the Weyl compensated volume preserving diffeomorphism and the conventional unimodular gravity discussed in the literature (see e.g. \cite{Padilla:2014yea} for a recent review) because both theories retain the volume preserving diffeomorphism as (a part of) the symmetry of the action and the deceptive similarity may cause a confusion. 
After all, the conventional unimodular gravity is nothing but the gauge fixed form of Einstein gravity, and therefore the Einstein-Hilbert term is allowed. Our gravity is different because although the Einstein-Hilbert term is invariant under the diffeomorphism (with or without volume preserving condition),  we have to supplement the Weyl transformation to preserve the unimodular gravity gauge condition in our case if the diffeomorphism  transformation is not volume preserving. Since the compensating Weyl transformation is not a symmetry of the Einstein-Hilbert action, our Virial supergravity does not admit the Einstein-Hilbert term (without couplings to matter).

To illustrate the use of the Virial connection $G_\mu$, we revisit the coupling of the linearized Virial supergravity field and the Virial supercurrent supermultiplet

\begin{align}\int d^4x d^4 \theta \left( J^V_{\alpha \dot{\alpha}} H^{\alpha \dot{\alpha}}  +6 UO  \right) \ .  
\end{align}
Here the $J^V_{\alpha \dot{\alpha}}$ is the Virial supercurrent supermultiplet satisfying the superspace conservation equations \eqref{defv}, and $O$ is the potential for the chiral superfield $\eta_{\alpha}$ in \eqref{defv} such that $\eta_{\alpha} = -\frac{1}{2}\bar{D}^2\bar{D}_{\alpha} O$. 

By using the component expressions discussed in this section, we find the coupling 
\begin{align}
\int d^4x \left(-2 A^\mu j_\mu  + 2h^{\mu\nu} T_{\mu\nu} + 2G^\mu B_\mu + \cdots \right) \cr
= \int d^4x \left( -2A^\mu j_\mu + h^{\mu\nu} (T_{\mu\nu} + T_{\nu\mu}) + 2G^\mu B_\mu + \cdots \right)\ 
\end{align}
among others in the Wess-Zumino gauge supplemented with the unimodular gravity condition. 
The linearized metric $h_{\mu\nu}$ is symmetric and traceless.
We emphasize again that the symmetrized energy-momentum tensor $T_{\mu\nu} + T_{\nu\mu}$ is not conserved by itself (i.e. $\partial^\mu (T_{\mu\nu} + T_{\nu\mu}) \neq 0$). We also note that the divergence free vector field $G^\mu$ couples to the Virial current $B_\mu$.
By recalling that the anti-symmetric part of the energy-momentum tensor $T_{\mu\nu}$ is given by the rotation of the Virial current, the non-invariance can be cast into the form
\begin{align}
\int d^4x 2(\Gamma^\mu +G^\mu)B_\mu + \text{diff inv} \ . \label{connection}
\end{align}
The diffeomorphism non-covariance of the Christoffel connection $\Gamma^\mu$ is precisely cancelled by the non-tensorial transformation of $G^\mu$ under the Weyl compensated volume preserving diffeomorphism. In the unimodular gravity gauge $\det{g} = -1$, the expression \eqref{connection} would be valid beyond the linearized order (but see below at the end of this section).

We would like to discuss some examples.
In section 3, we have discussed the coupling of the Virial supergravity to a (free) chiral superfield $\chi$ with the Weyl weight $-1+i\alpha$. When $\alpha = 0$, the action is super Weyl invariant and the component action is
\begin{align}
\int d^4x \sqrt{-g} \left( \partial^\mu \bar{\chi}\partial_\mu \chi + \frac{1}{6} R \bar{\chi}\chi + \frac{i}{3} A_\mu (\bar{\chi} \partial^\mu \chi - \chi\partial^\mu \bar{\chi}) + \mathrm{fermions} \ \right) .  
\end{align}
We may fix the residual super Weyl gauge symmetry by fixing $\det{g} = -1$. 

At the first order in $\alpha$ (as well as the first order in non-linearity of the Virial supergravity fields), the effects of non-zero $\alpha$ in the unimodular gravity gauge are given by 
\begin{align}
\alpha \int d^4x \left(i(\Gamma_\mu + G_\mu)(\bar{\chi} \partial^\mu \chi - \chi\partial^\mu \bar{\chi})  +    A_\mu \partial^\mu (\bar{\chi}\chi) + \mathrm{fermions} \right) \ .
\end{align}
As argued before, the variation with respect to the metric gives the symmetric but non-conserved  energy-momentum tensor due to the non-covariance of the Christoffel connection $\Gamma_\mu$. The conservation equation, however, will be compensated by the anti-symmetric part from the variation of $G_\mu$ (or its potential $B_{\mu\nu}$).

The next example is the higher derivative kinetic terms  \eqref{virihh} for the pure Virial supergravity. The Weyl invariant term (proportional to $\alpha$ in \eqref{virihh}) gives the usual conformal supergravity action
\begin{align}
\int d^4x \sqrt{-g} \left(W_{\mu\nu\rho\sigma} W^{\mu\nu\rho\sigma} - \frac{1}{4}F_{\mu\nu} F^{\mu\nu} + \mathrm{fermions} \right) \ ,
\end{align}
where $W_{\mu\nu\rho\sigma}$ is the Weyl tensor and $F_{\mu\nu} = \partial_\mu A_\nu -\partial_\nu A_\mu$ is the field strength of $A_\mu$. This action is ubiquitous in any formulation of the supergravity because it does not involve any compensators. What is unique in the Virial supergravity is the non Weyl invariant (but still scale invariant) term proportional to $\beta$ in \eqref{virihh}.

The corresponding linearized action is
\begin{align}
\int d^4x d^2\theta \mathcal{W}^{\alpha} \mathcal{W}_{\alpha} + \mathrm{c.c} \ , \label{linearizedr2}
\end{align}
with $\mathcal{W}_{\alpha} = \bar{D}^2 D_{\alpha} (\partial^\mu H_\mu -6U)$. 
In components (in the unimodular gravity gauge $h=0$), it reads\footnote{If the coupling constant $\beta$ is complex, it also gives the ``topological term" out of $\Gamma_\mu +G_\mu$.}
\begin{align}
\int d^4x \left( \frac{1}{4}\left(\partial_\mu (\Gamma_\nu+G_\nu)-\partial_\nu(\Gamma_\mu+G_\mu) \right)^2 -2(\partial^\mu A_\mu)^2 + \mathrm{fermions} \right) \ . \label{actionr}
\end{align}

It is instructive to compare the action \eqref{actionr} with the similar higher derivative action in the new-minimal supergravity that is obtained by replacing  $\mathcal{W}_{\alpha}$ in \eqref{linearizedr2} with $\mathcal{W}_{\alpha} = \bar{D}^2 D_{\alpha} {([D_{\beta},\bar{D}_{\dot{\beta}}] H^{\beta \dot{\beta}} - 3L)}$ constructed out of the linearized new minimal supergravity compensator $L$ \cite{Cecotti:1987qe}. In the Wess-Zumino gauge, we have
\begin{align}
S_{\mathrm{new-minimal}}^{R^2} = \int d^4x \sqrt{-g}\left( -R^2 + \frac{1}{4}\left(\partial_\mu (A_\nu+\mathcal{G}_\nu)-\partial_\nu(A_\mu+\mathcal{G}_\mu) \right)^2+ \mathrm{fermions} \right) \ , \label{nactionr}
\end{align}
where $\mathcal{G}_\mu$ in $L \sim ({\theta}\sigma^\mu \bar{\theta}) \mathcal{G}_\mu$ is a divergence free vector field as our $G_\mu$ in $V$.
This higher derivative new-minimal supergravity action is diffeomorphism invariant in the usual sense, and the gauge field $A_\mu$ acquires the {\it gauge invariant} kinetic term. This gauge invariance is necessary because it transforms as $A_\mu \to A_\mu + \partial_\mu g$ under the non-compensated R-symmetry gauging of the new minimal supergravity.

In contrast, in our Virial supergravity, we do not obtain the $R^2$ term (due to extra ``$i$"s in the supercurrent supermultiplets).
Indeed, the pure $R^2$ term is not invariant under the Weyl compensated volume preserving diffeomorphism.  Rather the gravity action is given by the gauge invariant Maxwell form of the (Virial) connection $\Gamma_\mu + G_\mu$. 
 Again note that the non-covariance of the (linearized) Christoffel connection in \eqref{actionr} is cancelled by the variation of our Virial connection $G_\mu$. As for the vector field $A_\mu$, the action $(\partial^\mu A_\mu)^2$ looks like the ``gauge fixing term". Since the R-symmetry is compensated in the Virial supergravity, there is no necessity of the gauge invariance for the $A_\mu$ action in the Wess-Zumino gauge, and the seemingly gauge non-invariant term $(\partial^\mu A_\mu)^2$ is allowed.

To conclude the section, let us briefly discuss the nature of the unitarity constraint on our superconformal compensator $V$ at the non-linear order. Finding a solution of the constraint quickly becomes cumbersome beyond the first order. Even at the zeroth order in the background vierbein superfield $H_a$, the non-linear constraint
\begin{align}
\gamma \bar{\gamma} = 1 \cr
\bar{D}^2 \gamma = 0 \ 
\end{align}
is highly non-trivial. Focusing on the bosonic sector only, we find
\begin{align}
\gamma = e^{i\phi} + i(\theta \sigma^\mu \bar{\theta}) e^{i\phi} G_\mu + \theta^2\bar{\theta}^2 \left(\frac{1}{2}\partial_\mu (e^{i\phi} G^\mu) -\frac{1}{4}\partial^\mu \partial_\mu e^{i\phi} \right) +  \mathrm{fermions} \ ,
\end{align}
where $\phi$ and $G_\mu$ are real fields.
In the linearized approximation, $G_\mu$ was divergence free $\partial^\mu G_\mu =0$, which may be locally solved by introducing the two-form potential $B_{\mu\nu}$ as $G_\mu = \frac{1}{2}\epsilon_{\mu\nu\rho\sigma} \partial^\nu B^{\rho\sigma}$. However, the non-linear constraint is
\begin{align}
\partial_\mu G^\mu = \frac{1}{2}G_\mu G^\mu - \frac{1}{2}(\partial_\mu\phi \partial^\mu \phi) + \mathrm{fermions} \ , 
\end{align}  
so there is no immediate solution available. We may use the superconformal gauge freedom to fix $\phi = 0$ or any other values, but still the constraint is non-linear.
We therefore conclude that pursuing the component expression by solving all the constraints is not the best way to present the physical contents of the Virial supergravity beyond the linearized order, but this does not devalue the full non-linear superfield expressions in section 3.

Alternatively one may get rid of the non-linear unitarity constraint completely by introducing the Lagrange multiplier real superfield $\Lambda$. The action
\begin{align}
\int d^8z E^{-1} \Lambda(V\bar{V} -1)
\end{align}
for an unconstrained covariantly linear {\it complex} superfield $V$ effectively gives rise to the Virial supergravity. Here under the super Weyl transformation $ \Lambda \to e^{-\sigma -\bar{\sigma}} \Lambda $ and $V \to e^{\sigma-\bar{\sigma}} V$. 
It is equivalent to the $n\to \infty$ limit of the non-minimal supergravity coupled to a particular real superfield $\Lambda$. The magnitude of the $\theta$ independent components in $V$ can be identified with the determinant of the metric in the Wess-Zumino gauge, and then the Lagrange multiplier gives the unimodular condition. The constraint in the component expansions becomes linear, but of course, the equations of motions are non-linear.
It would be interesting to find a connection to the unimodular supergravity studied in \cite{Nishino:2001gd}.

\section{Discussions}

In this paper, we have constructed a novel type of $\mathcal{N}=1$ supergravity  from gauging the Virial supercurrent supermultiplet in $d=1+3$ dimensions. Unlike the new minimal supergravity based on the R-current supermultiplet, the resulting supergravity is not equivalent to the old minimal supergravity. The emerging geometric picture is very peculiar (e.g. non-geometrical connection, unimodular condition etc), and it would be interesting to formulate the Virial supergravity purely in terms of the supergeometry by directly working in the superconformal gauge $\varphi = 1$ rather than relying on the superconformal embedding with the help of the  other density type compensators.

Given the classification of the linearized supersymmetric massless spin two actions in \cite{Gates:2003cz}, we declare that we have completed the classification of the irreducible supergravities parameterized by a complex number $n$ at the full non-linear level. We have three minimal supergravities, the old minimal supergravity ($n=-\frac{1}{3}$ with the covariantly chiral compensator), the new minimal supergravity ($n=0$ with the covariantly linear real compensator) and the Virial supergravity ($n= \infty$ with the covariantly linear unitary compensator).  With this respect, the moduli space of the irreducible  supergravities parameterized by $n$ may be better treated as compact $\mathbb{CP}^1$ with four  special points, out of which the two points ($n=\pm\frac{1}{3}$) may be identified.\footnote{When we look at the Weyl transformation of the linear compensator $\Psi \to \Psi' = \exp(-d_{(+)}\sigma -\bar{\sigma}) \Psi = \exp(\frac{3n-1}{3n+1}\sigma - \bar{\sigma}) \Psi$, we realize $d_{+} = \pm1, 0,\infty$ are special points where the compensator can be reduced. In addition to the three points we have discussed, we have the fourth point $d_{+}=0$ or $n=\frac{1}{3}$, at which we can {\it replace} the linear condition with the anti-chiral condition. It then
 corresponds to the covariantly anti-chiral compensator, which is equivalent to the covariantly chiral compensator with $n=-\frac{1}{3}$ by the complex conjugation. We would like to W.~Linch for the fruitful discussions.} The other values of $n$ yield non-minimal supergravities. 

What would be the use of this Virial supergravity? One possibility is the application to the localization computations in rigid supersymmetric field theories in curved space-time.  In the literature, they have used the new minimal supergravity to obtain rigid supersymmetric field theories in curved space-time \cite{Festuccia:2011ws}\cite{Klare:2012gn}\cite{Liu:2012bi}\cite{Kuzenko:2012vd}\cite{Cassani:2013dba}\cite{Assel} and have discussed the localization computations of partition functions or various correlation functions \cite{Closset:2013vra}\cite{Closset:2014uda}. The new minimal supergravity allows different choices of the R-symmetry, and the choice of the background supergravity fields lead to non-trivial additions to the supersymmetric indices and partition functions. In a similar way, our Virial supergravity accommodates a different choice of dilatation currents, and the background gauging from our non-geometric connection may give a novel way to place the supersymmetric field theories on a curved manifold. 

In order to put the supersymmetric field theories on a generic curved manifold, there is a folklore that we need the R-symmetry (in addition to the $\mathrm{Spin}^{\mathbb{C}}$ structure) to ensure the well-defined supercharge. Our Virial supergravity can be formulated without the R-symmetry, and it may be possible to substitute the dilatation symmetry for the R-symmetry to obtain the rigid supersymmetric field theories for this purpose. We recall that the arguments given in \cite{Closset:2013vra}\cite{Closset:2014uda} do not exclude such a possibility.
We would like to come back to this question in the future.

On the other hand, it seems less likely that our Virial supergravity has phenomenological applications. The Virial supergravity is too predictive in the sense that matter must be scale invariant. Nevertheless, it may not be completely excluded because our universe may reside in the spontaneously broken phase of scale invariance (with the very weakly coupled dilaton).\footnote{This viewpoint is slightly idiosyncratic because it is more natural to regard the dilaton multiplet as the superconformal compensator then.}
 The higher derivative kinetic terms discussed in our paper may have applications to cosmology similarly to the $R^2$ term in the new-minimal supergravity.

All the discussions in this paper are purely classical, but it is imperative to understand the quantum aspects of the Virial supergravity. In many supersymmetric quantum field theories, the classical scale invariance, which is the basis of our construction of the Virial supergravity, is broken, and the consistent matter couplings are more constrained. This was also the case for the new minimal supergravity in which the R-symmetry may become anomalous. To avoid confusion, we emphasize that the superconformal compensator approach itself does not break down from the scale and R-symmetry anomaly. What breaks down is the possibility to compensate the anomaly with a given set of compensators.

We hope that both the Venus physicists and Earth physicists agree that the Virial supergravity is not an imaginary supergravity.

\section*{Acknowledgements}
This work is supported by Sherman Fairchild Senior Research Fellowship at California Institute of Technology  and DOE grant number DE-SC0011632.
The author would like to thank S.~Deser for amusing conversations on the Venus physicist approach.
He also thanks S.~Kuzenko and  W.~D.~Linch for the correspondence and discussions. They have kindly read the manuscript and given various suggestions on it.


\begin{thebibliography}{99}


\bibitem{Feynman:1996kb} 
  R.~P.~Feynman, F.~B.~Morinigo, W.~G.~Wagner and B.~Hatfield,
  Reading, USA: Addison-Wesley (1995) 232 p. (The advanced book program)


\bibitem{Deser:1969wk} 
  S.~Deser,
  Gen.\ Rel.\ Grav.\  {\bf 1}, 9 (1970)
  [gr-qc/0411023].

\bibitem{Deser:1979zb} 
  S.~Deser, J.~H.~Kay and D.~G.~Boulware,
  Physica {\bf 96A}, 141 (1979).

\bibitem{Gates:1983nr} 
  S.~J.~Gates, M.~T.~Grisaru, M.~Rocek and W.~Siegel,
  hep-th/0108200.

\bibitem{Buchbinder:1998qv} 
  I.~L.~Buchbinder and S.~M.~Kuzenko,
  Bristol, UK: IOP (1998) 656 p


\bibitem{Ferrara:1974pz} 
  S.~Ferrara and B.~Zumino,
  Nucl.\ Phys.\ B {\bf 87}, 207 (1975).



\bibitem{Wess:1978bu} 
  J.~Wess and B.~Zumino,
  Phys.\ Lett.\ B {\bf 74}, 51 (1978).

\bibitem{Stelle:1978ye} 
  K.~S.~Stelle and P.~C.~West,
  Phys.\ Lett.\ B {\bf 74}, 330 (1978).

\bibitem{Ferrara:1978em} 
  S.~Ferrara and P.~van Nieuwenhuizen,
  Phys.\ Lett.\ B {\bf 74}, 333 (1978).










\bibitem{Akulov:1976ck} 
  V.~P.~Akulov, D.~V.~Volkov and V.~A.~Soroka,
  Theor.\ Math.\ Phys.\  {\bf 31}, 285 (1977)
  [Teor.\ Mat.\ Fiz.\  {\bf 31}, 12 (1977)].


\bibitem{Sohnius:1981tp} 
  M.~F.~Sohnius and P.~C.~West,
  Phys.\ Lett.\ B {\bf 105}, 353 (1981).







\bibitem{Gates:1981yc} 
  S.~J.~Gates, Jr., M.~T.~Grisaru and W.~Siegel,
  Nucl.\ Phys.\ B {\bf 203}, 189 (1982).




\bibitem{Magro:2001aj} 
  M.~Magro, I.~Sachs and S.~Wolf,
  Annals Phys.\  {\bf 298}, 123 (2002)
  [hep-th/0110131].



\bibitem{Komargodski:2010rb} 
  Z.~Komargodski and N.~Seiberg,
  JHEP {\bf 1007}, 017 (2010)
  [arXiv:1002.2228 [hep-th]].



\bibitem{Dumitrescu:2011iu} 
  T.~T.~Dumitrescu and N.~Seiberg,
  JHEP {\bf 1107}, 095 (2011)
  [arXiv:1106.0031 [hep-th]].


\bibitem{Kuzenko:2011rd} 
  S.~M.~Kuzenko and G.~Tartaglino-Mazzucchelli,
  JHEP {\bf 1112}, 052 (2011)
  [arXiv:1109.0496 [hep-th]].





\bibitem{Gates:2003cz} 
  S.~J.~Gates, Jr.., S.~M.~Kuzenko and J.~Phillips,
  Phys.\ Lett.\ B {\bf 576}, 97 (2003)
  [hep-th/0306288].

\bibitem{Buchbinder:2002gh} 
  I.~L.~Buchbinder, S.~J.~Gates, Jr., W.~D.~Linch, III and J.~Phillips,
  Phys.\ Lett.\ B {\bf 535}, 280 (2002)
  [hep-th/0201096].



\bibitem{Kuzenko:2010am} 
  S.~M.~Kuzenko,
  JHEP {\bf 1004}, 022 (2010)
  [arXiv:1002.4932 [hep-th]].


\bibitem{Zheng:2010xx} 
  S.~Zheng and J.~h.~Huang,
  Class.\ Quant.\ Grav.\  {\bf 28}, 075012 (2011)
  [arXiv:1007.3092 [hep-th]].

\bibitem{Kuzenko:2010ni} 
  S.~M.~Kuzenko,
  Eur.\ Phys.\ J.\ C {\bf 71}, 1513 (2011)
  [arXiv:1008.1877 [hep-th]].

\bibitem{Nakayama:2012nd} 
  Y.~Nakayama,
  Phys.\ Rev.\ D {\bf 87}, 085005 (2013)
  [arXiv:1208.4726 [hep-th]].

\bibitem{Siegel:1978mj} 
  W.~Siegel and S.~J.~Gates, Jr.,
  Nucl.\ Phys.\ B {\bf 147}, 77 (1979).




\bibitem{Festuccia:2011ws} 
  G.~Festuccia and N.~Seiberg,
  JHEP {\bf 1106}, 114 (2011)
  [arXiv:1105.0689 [hep-th]].

\bibitem{Dumitrescu:2012ha} 
  T.~T.~Dumitrescu, G.~Festuccia and N.~Seiberg,
  JHEP {\bf 1208}, 141 (2012)
  [arXiv:1205.1115 [hep-th]].


\bibitem{Nakayama:2013coa} 
  Y.~Nakayama,
  JHEP {\bf 1308}, 049 (2013)
  [arXiv:1305.2937, arXiv:1305.2937 [hep-th]].

\bibitem{Maldacena:2011mk} 
  J.~Maldacena,
  arXiv:1105.5632 [hep-th].



\bibitem{Dienes:2009td} 
  K.~R.~Dienes and B.~Thomas,
  Phys.\ Rev.\ D {\bf 81}, 065023 (2010)
  [arXiv:0911.0677 [hep-th]].









\bibitem{Nakayama:2013is} 
  Y.~Nakayama,
  arXiv:1302.0884 [hep-th].


\bibitem{Breitenlohner:1977jn} 
  P.~Breitenlohner,
  Nucl.\ Phys.\ B {\bf 124}, 500 (1977).
\bibitem{Brink:1978iv} 
  L.~Brink, M.~Gell-Mann, P.~Ramond and J.~H.~Schwarz,
  Phys.\ Lett.\ B {\bf 74}, 336 (1978).

\bibitem{Fradkin:1985am} 
  E.~S.~Fradkin and A.~A.~Tseytlin,
  Phys.\ Rept.\  {\bf 119}, 233 (1985).

\bibitem{Cecotti:1987qe} 
  S.~Cecotti, S.~Ferrara, M.~Porrati and S.~Sabharwal,
  Nucl.\ Phys.\ B {\bf 306}, 160 (1988).


\bibitem{Riva:2005gd} 
  V.~Riva and J.~L.~Cardy,
  Phys.\ Lett.\ B {\bf 622}, 339 (2005)
  [hep-th/0504197].

\bibitem{ElShowk:2011gz} 
  S.~El-Showk, Y.~Nakayama and S.~Rychkov,
  Nucl.\ Phys.\ B {\bf 848}, 578 (2011)
  [arXiv:1101.5385 [hep-th]].

\bibitem{Padilla:2014yea} 
  A.~Padilla and I.~D.~Saltas,
  arXiv:1409.3573 [gr-qc].


\bibitem{Nishino:2001gd} 
  H.~Nishino and S.~Rajpoot,
  Phys.\ Lett.\ B {\bf 528}, 259 (2002)
  [hep-th/0107202].


\bibitem{Klare:2012gn} 
  C.~Klare, A.~Tomasiello and A.~Zaffaroni,
  JHEP {\bf 1208}, 061 (2012)
  [arXiv:1205.1062 [hep-th]].

\bibitem{Liu:2012bi} 
  J.~T.~Liu, L.~A.~Pando Zayas and D.~Reichmann,
  JHEP {\bf 1210}, 034 (2012)
  [arXiv:1207.2785 [hep-th]].



\bibitem{Kuzenko:2012vd} 
  S.~M.~Kuzenko,
  JHEP {\bf 1303}, 024 (2013)
  [arXiv:1212.6179 [hep-th]].


\bibitem{Cassani:2013dba} 
  D.~Cassani and D.~Martelli,
  JHEP {\bf 1310}, 025 (2013)
  [arXiv:1307.6567 [hep-th]].

\bibitem{Assel} 
  B.~Assel, D.~Cassani and D.~Martelli,
  arXiv:1410.6487 [hep-th].


\bibitem{Closset:2013vra} 
  C.~Closset, T.~T.~Dumitrescu, G.~Festuccia and Z.~Komargodski,
  JHEP {\bf 1401}, 124 (2014)
  [arXiv:1309.5876 [hep-th]].

\bibitem{Closset:2014uda} 
  C.~Closset, T.~T.~Dumitrescu, G.~Festuccia and Z.~Komargodski,
  Phys.\ Rev.\ D {\bf 90}, 085006 (2014)
  [arXiv:1407.2598 [hep-th]].

\end{thebibliography}
\end{document}